\documentclass[fleqn,12pt,twoside]{article}
\usepackage{espcrc1}
\usepackage{graphicx}
\title{Chiral dynamics and  pionic 1s states of Pb and Sn
  isotopes\thanks{Work supproted in part by BMBF and GSI; talk
  presented by W.~Weise at PANIC'02 }}

\author{E.E.~Kolomeitsev\address[ECT]{ECT*, Villa Tambosi, I-38050 Villazzano
(Trento),Italy}\address[NBI]{The Niels Bohr Institute, DK-2100 Copenhagen,
Denmark},
N.~Kaiser\address[TUM]{Physik-Department, TU M\"unchen, D-85747 Garching,
Germany},
W.~Weise\addressmark[ECT]\addressmark[TUM]
}
\begin{document}

\maketitle


Recent accurate data on $1s$ states of $\pi^-$ bound to
Pb~\cite{gilg} and Sn~\cite{suzuki} isotopes have set new standards
and constraints for the  detailed analysis of s-wave pion-nucleon
interactions. This topic has a long history~\cite{ew} culminating
in various attempts to understand the notorious "missing
repulsion" in the $\pi$-nucleus interaction: the standard ansatz
for the  (energy independent) s-wave pion-nucleus optical
potential, given in terms of the empirical threshold $\pi N $
amplitudes times densities $\rho_{p,n}$ and supplemented by sizable
double-scattering corrections, still misses the observed repulsive
interaction by a large amount. This problem has
traditionally been circumvented on
purely phenomenological grounds by introducing an extraordinarily
large repulsive real part (${\rm Re} B_0$) in the $\rho^2$ terms
of the $\pi$-nucleus potential. The  arbitrariness  of this
procedure is of course unsatisfactory.

In recent papers~\cite{kkw,kw} we have re-investigated this issue
from the point of view of the distinct explicit energy dependence
of the pion-nuclear polarization operator~\cite{kkw} in a
calculation based on systematic in-medium chiral perturbation
theory~\cite{kw,wom}. Ref.~\cite{kkw} has also clarified the
relationship to a working hypothesis launched
previously~\cite{w,ky}: that the extra repulsion needed in the
s-wave pion-nucleus optical  potential at least partially reflects
the tendency toward chiral symmetry restoration in a dense medium.
To leading order, this information is encoded in the in-medium
reduction of the pion decay constant $f_\pi$, which, by its inverse
square, drives the isospin-odd pion-nucleon amplitudes close to
threshold. The aim of the this note is to present an updated
summary of the situation and to compare with the new  Sn
data~\cite{suzuki}. A detailed assessment of the overall
systematics covering the complete pionic atoms data base has
recently been given in ref.~\cite{f}, using optical potential
phenomenology.

The starting point is the energy- and  momentum-dependent
polarization operator (the pion  self-energy) $\Pi(\omega,\vec
q;\rho_p,\rho_n)$\,. In the limit of very low proton and neutron
densities, $\rho_{p,n}$, the pion self-energy reduces to
$\Pi=-(T^+\,\rho+T^-\, \delta\rho)$ with $\rho=\rho_p+\rho_n$ and
$\delta \rho=\rho_p-\rho_n$, where $T^{\pm}$ are the isospin-even
and isospin-odd off-shell $\pi N$ amplitudes. In the
long-wavelength limit ($\vec{q}\to 0$), chiral symmetry (the
Weinberg-Tomozawa theorem) implies $T^-(\omega)=\omega/(2\,
f_\pi^2)+\mathcal{O}(\omega^3)$\,. Together with the observed
approximate vanishing of the isospin-even threshold amplitude
$T^+(\omega=m_\pi)$, it is clear that $1s$  states of pions bound to
heavy, neutron rich nuclei are a particularly sensitive source of
information for in-medium chiral dynamics.

At the same time, it has
long been known that term of  non-leading order in density (double
scattering (Pauli) corrections of order $\rho^{4/3}$, absorption
effects of order $\rho^2$ etc.) are important. The aim must,
therefore, be to arrive at a consistent expansion of the pion
self-energy in powers of the Fermi momentum $k_{\rm F}$ together
with the chiral low-energy expansion in  $\omega, |\vec{q}\, |$
and $m_\pi$\,.  In-medium chiral effective field theory provides a
framework for this approach. We apply it here systematically up to
two-loop order,  following ref.~\cite{kw}. Double scattering
corrections are fully incorporated at this order. Absorption
effects and corresponding dispersive corrections appear  at the
three-loop level and through  short-distance dynamics
parameterized by $\pi N N$ contact terms, not explicitly calculable
within the effective low-energy theory. The imaginary parts
associated with these terms are well constrained  by the
systematics  of observed  widths of pionic atom levels throughout
the periodic table. (We use ${\rm Im} B_0=-0.063 m_\pi^4$ in the
s-wave absorption term, $\Delta \Pi^{\rm abs}_{\rm S}=-8\, \pi\,
(1+m_\pi/2 M)\, B_0\, \rho_p\, (\rho_p+\rho_n)$, and the canonical
parameterization  of p-wave parts, in accordance with
refs.~\cite{ew,f}). The real part of $B_0$ is still the primary
source of theoretical uncertainty. In practice, our strategy is to
start from ${\rm Re} B_0=0$ (as suggested also by the detailed
analysis of the pion-deuteron scattering length) and then discuss
the possible error band  induced by varying ${\rm B_0}$ within
reasonable limits~\cite{kkw}.

We proceed by using the local density approximation (with gradient
expansion for p-wave interactions, $\vec{q\,}^2\, F(\rho)\to
\vec{\nabla}\, F(\rho(\vec r\, ))\, \vec{\nabla}$) and solve the
Klein-Gordon equation
\begin{eqnarray}\label{kge}
\Big[ \Big(\omega-V_c(\vec r\,)\Big)^2+\vec\nabla^2-m_\pi^2-
\Pi\Big(\omega-V_c(\vec r\,);
\rho_p(\vec r\, ),\rho_n(\vec r\,)\Big)\Big]\phi(\vec r\, )=0\,.
\end{eqnarray}
Note that the explicit energy dependence of $\Pi$ requires that
the Coulomb potential $V_c(\vec r\, )$ must be introduced in the
canonical gauge invariant way wherever the pion energy $\omega$
appears. This  is an important feature that has generally been
disregarded in previous analysis.

\begin{figure}[t]
\begin{minipage}[t]{77mm}
\begin{center}
\parbox{6.4cm}{
\includegraphics[width=6.4cm,clip=true]{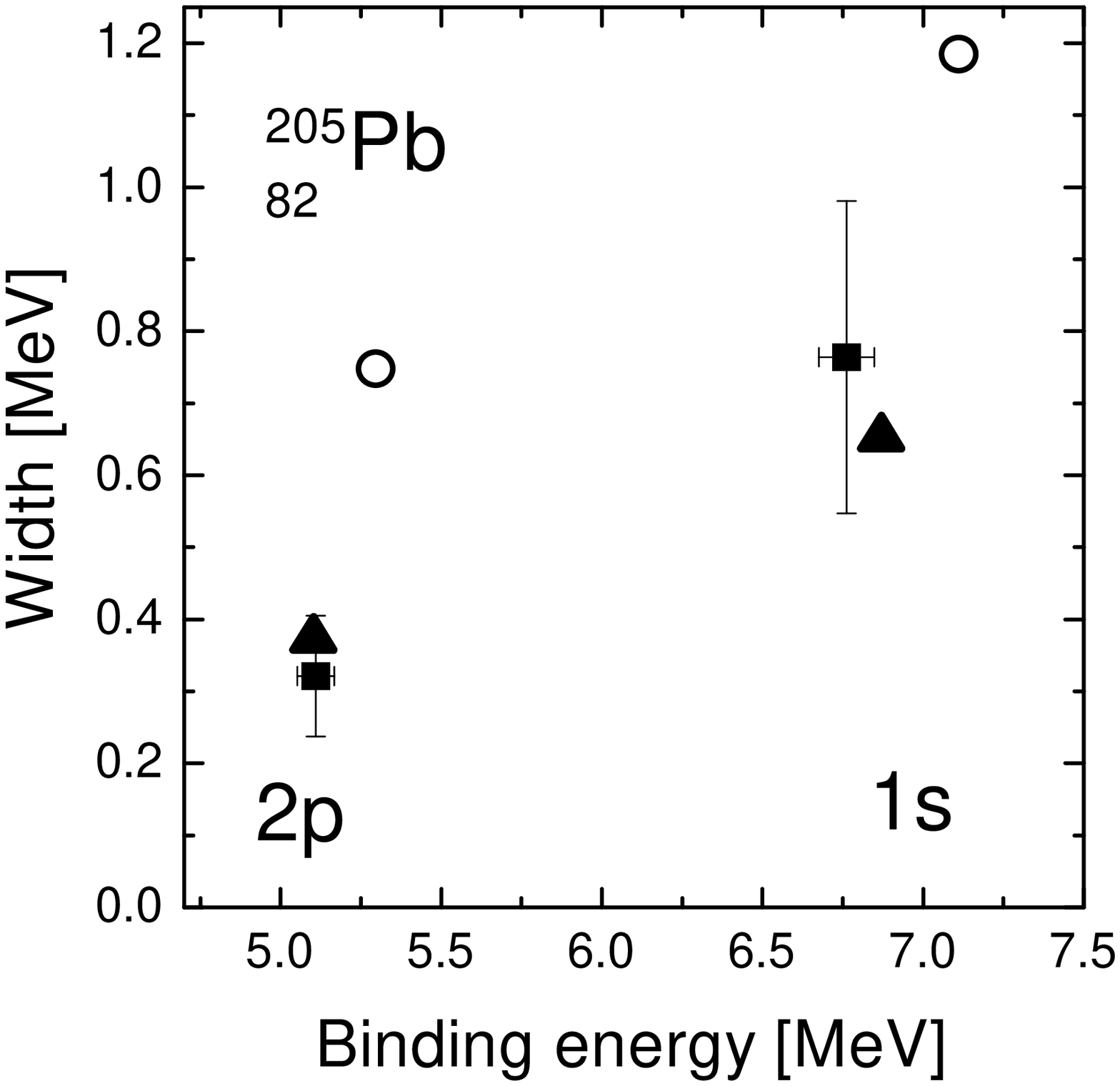}}
\end{center}
\caption{Binding energies and widths of pionic $1s$ and $2p$ states in
  $^{205}$Pb. Experimental data from~\cite{gilg}. Full  triangles:
  results of two-loop in-medium chiral perturbation theory, keeping
  the explicit energy dependence in the s-wave polarization
  operator. Open circles: energy independent potential as described in
  text (see ref.~\cite{kkw} for details). Note that ${\rm Re}B_0=0$ in
  both cases.}
\label{fig:pb}
\end{minipage}
\hspace{\fill}
\begin{minipage}[t]{77mm}
\begin{center}
\parbox{6.8cm}{
\includegraphics[width=6.8cm,clip=true]{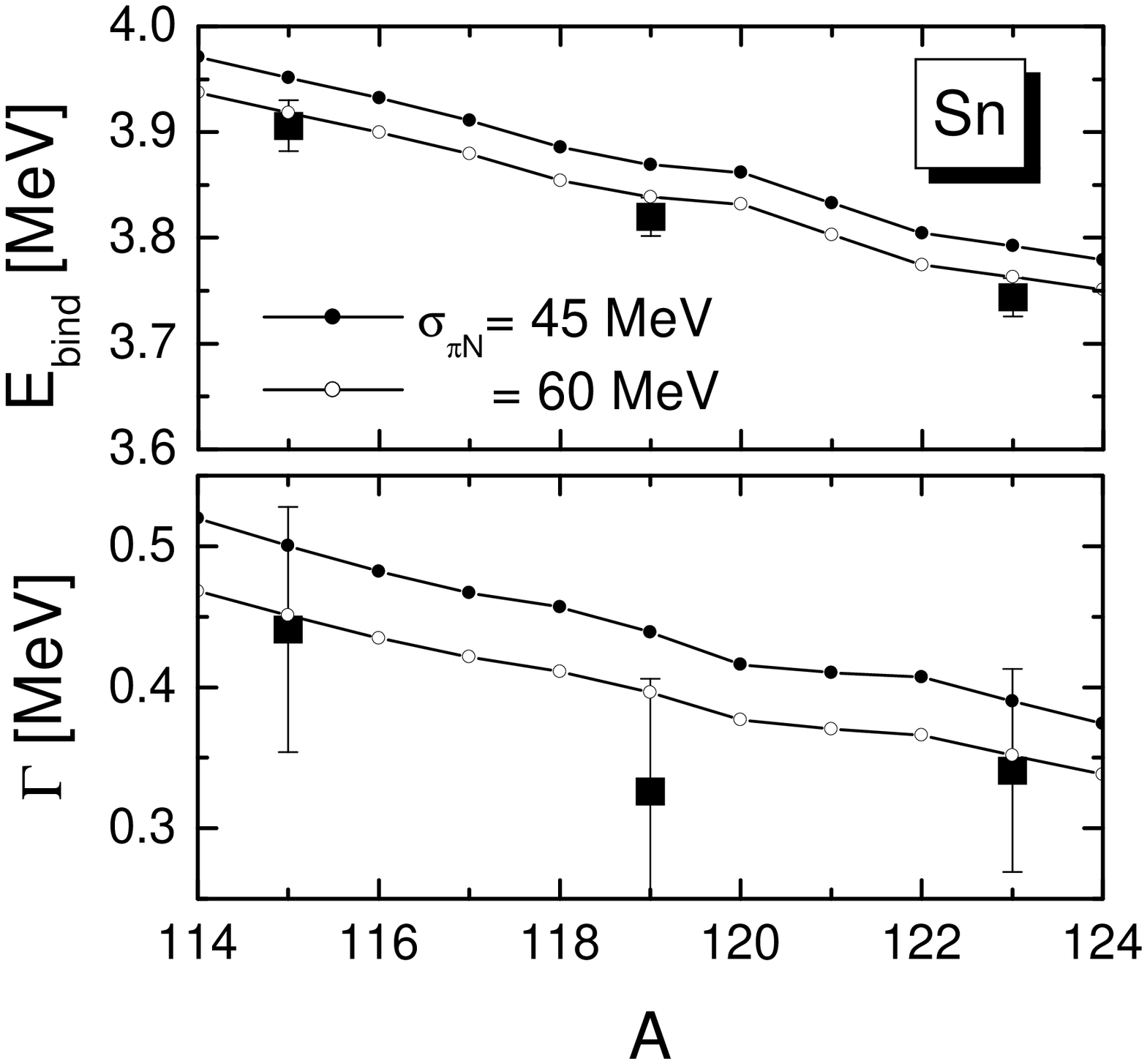}}
\end{center}
\caption{
Binding energies, $E_{\rm bind}$, and widths, $\Gamma$, of pionic $1s$ states in Sn isotopes. The
curves show predictions~\cite{kkw03} based on the explicitly  energy
dependent pionic s-wave polarization operator calculated in two-loop
in-medium chiral perturbation theory~\cite{kkw}. The sensitivity to
the $\pi N$ sigma term (input) is also shown. Data from ref.~\cite{suzuki}.
}
\label{fig:sn}
\end{minipage}
\end{figure}

With input specified in details in ref.~\cite{kkw}, we have solved
eq.~(\ref{kge}) with the explicitly energy dependent pion
self-energy, obtained in two-loop in-medium chiral perturbation
theory for the s-wave part, adding the time-honored
phenomenological p-wave piece. The results for the binding energies
and widths of $1s$ and $2p$ states in pionic $^{205}$Pb are shown
in Fig.~\ref{fig:pb} (triangles). Also shown for comparison is the
outcome of a calculations using a "standard" phenomenological
(energy independent) s-wave optical potential,
\begin{eqnarray}\label{pis}
\Pi_{\rm S}=-T_{\rm eff}^+\, \rho-T_0^-\,\delta
\rho+\Delta\Pi_{\rm S}^{\rm abs}\,,
\end{eqnarray}
with  $T_{\rm eff}^+=T_0^+-\frac{3 k_{\rm F}(\vec r)}{8\, \pi^2}
\, \Big[(T_0^+)^2+2\, (T_0^-)^2\Big]$ and the amplitudes
$T_0^{\pm}\equiv T^{\pm}(\omega=m_\pi)$ taken fixed at their
threshold values. This approach fails and shows the "missing
repulsion" syndrome, leading  to a substantial overestimate of the
widths. Evidently, a mechanism is needed to reduce the overlap of
the bound pion wave functions with the nuclear  density
distributions, The explicit energy  dependence in $T^{\pm}$
provides such a mechanism: the replacement
$\omega \to \omega-V_c(\vec r\,)>m_\pi$ increases the repulsion in $T^-$
and  disbalances the "accidental" cancellation between the $\pi N$
sigma term $\sigma_{ N}$ and the range term proportional to
$\omega^2$ in $T^+$, such that $T^+(\omega-V_c)<0$ (repulsive).

Uncertainties in ${\rm Re} B_0$, in the radius and shape of the
neutron density distribution, and in the input for the sigma term
$\sigma_N$ have been analysed in ref.~\cite{kkw}. Their combined
effect falls within the experimental errors in Fig.~\ref{fig:pb}.

Using the same (explicitly energy dependent) scheme we have
predicted binding energies and widths for pionic $1s$ states bound
to a  chain of Sn isotopes. These calculations~\cite{kkw03}
include  a careful assessment of uncertainties in neutron
distributions. Results are shown in Fig.~\ref{fig:sn} in
comparison with experimental data~\cite{suzuki} reported at
PANIC'02 after the calculations. This figure also gives an
impression of the sensitivity with respect to variations of the
(input) $\pi N$ sigma term.

We now come to an important question of interpretation: do we
actually "observe" fingerprints of (partial) chiral symmetry
restoration in the high-precision data of deeply bound pionic
atoms with heavy nuclei, as anticipated in refs.~\cite{w,ky}? Is
this observation related to the "missing s-wave repulsion" that
has been recognized (but not resolved in a consistent  way) by
scanning the large amount of already existing pionic atom data?

To approach this question, recall that pionic  atom calculations
are traditionally done with \emph{energy-independent}
phenomenological optical potentials instead of explicitly energy
dependent pionic polarization functions. Let us examine the
connection between these two seemingly different approaches by
illustrating the leading-order driving mechanisms.

Consider a zero momentum pion in low density matter. Its energy
dependent leading-order polarization operator is $\Pi(\omega)=-
\Big[T^+(\omega)\, \rho+T^-(\omega)\, \delta\rho\Big]$\,, and the
in-medium dispersion equation at  $\vec {q}=0$ is
$\omega^2-m_\pi^2-\Pi(\omega)=0$\,. The chiral low-energy
expansion of the off-shell amplitudes $T^{\pm}(\omega)$ at $\vec {q}=0$
implies leading terms of the form:
\begin{eqnarray}
\label{t}
T^+(\omega)=\frac{\sigma_N-\beta\, \omega^2}{f_\pi^2}\,,
\quad T^-(\omega)=\frac{\omega}{2\, f_\pi^2}\,,
\end{eqnarray}
where $f_\pi=92.4$~MeV is the pion decay constant in vacuum and
$\sigma_N\simeq 0.05$~GeV.  The empirical $T^+(\omega=m_\pi)=
(-0.04\pm 0.09)~{\rm fm}\simeq 0$ sets the constraint $\beta\simeq
\sigma_N/m_\pi^2$\,.

Expanding $\Pi(\omega)$ around  the threshold, $\omega=m_\pi$\,,
we identify the commonly used effective (energy-independent)
s-wave optical potential $U_{\rm S}$ as:
\begin{eqnarray}\label{us}
2\, m_\pi\, U_{\rm S}=\frac{\Pi(\omega=m_\pi,\vec
q=0)}{1-\partial\Pi/\partial\omega^2}\,,
\end{eqnarray}
where $\partial\Pi/\partial\omega^2$ is taken at $\omega=m_\pi$\,.
Inserting (\ref{t}) and assuming $\delta\rho\ll \rho$ one finds:
\begin{eqnarray}
\label{us2}
U_{\rm S}\simeq -\frac{\delta \rho}{4\, f_\pi^2}\,
\left(1-\frac{\sigma_N\, \rho}{m_\pi^2\, f_\pi^2}\right)^{-1} =
-\frac{\delta \rho}{4\, f_\pi^{* 2}(\rho)}\,,
\end{eqnarray}
with the replacement $f_\pi\to f_\pi^*(\rho)$ of the pion decay
constant representing the in-medium wave function renormalization.
The expression (\ref{us2}) is just the one proposed previously in
ref.~\cite{w} on the basis of the relationship between the
in-medium changes of the chiral condensate $<\overline{q}\, q>$
and of the pion decay constant  associated with the time
component of the axial current. The explicitly energy dependent
chiral dynamics encoded in $\Pi(\omega)$ "knows" about these
renormalization effects. Their  translation into an equivalent,
energy-independent potential implies  $f_\pi\to f_\pi^*(\rho)$ as
given in eq.~(\ref{us2}). This statement holds to leading order.
Whether (important) higher order corrections permit a similar
interpretation needs still to be further explored.

\end{document}